\documentclass[fleqn]{llncs}
\usepackage{version}
\usepackage{tpis}

\title{Transmission Protocols  for Instruction Streams}
\author{J.A. Bergstra \and C.A. Middelburg}
\institute{Informatics Institute, Faculty of Science,
           University of Amsterdam, \\
           Science Park~107, 1098~XG Amsterdam, the Netherlands \\
           \email{J.A.Bergstra@uva.nl,C.A.Middelburg@uva.nl}}

\begin{document}

\maketitle

\begin{abstract}
Threads as considered in thread algebra model behaviours to be
controlled by some execution environment: upon each action performed by
a thread, a reply from its execution environment -- which takes the
action as an instruction to be processed -- determines how the thread
proceeds.
In this paper, we are concerned with the case where the execution
environment is remote: we describe and analyse some transmission
protocols for passing instructions from a thread to a remote execution
environment.
\begin{classcode}
D.2.1, D.2.4, F.1.1, F.3.1.
\end{classcode}
\end{abstract}

\section{Introduction}
\label{sect-intro}

The behaviours produced by sequential programs under execution are
behaviours to be controlled by some execution environment.
The execution environment concerned is increasingly more a remote
execution environment.
The objective of the current paper is to clarify the phenomenon of
remotely controlled program behaviours.

Basic thread algebra~\cite{BL02a}, \BTA\ in short, is a form of process
algebra tailored to the description and analysis of the behaviours
produced by sequential programs under execution.%
\footnote
{In~\cite{BL02a}, basic thread algebra is introduced under the name
 basic polarized process algebra.
}
Threads as considered in basic thread algebra model behaviours to be
controlled by some execution environment.
Threads proceed by performing steps, called basic actions in what
follows, in a sequential fashion.
The execution environment of a thread takes the basic actions performed
by the thread as instructions to be processed.
Upon each basic action performed by the thread, a reply from the
execution environment determines how the thread proceeds.
To achieve the objective of the current paper, we study some
transmission protocols for passing instructions from a thread to a
remote execution environment.

General process algebras, such as \ACP~\cite{BK84b,BW90},
CCS~\cite{HM85,Mil89} and CSP~\cite{BHR84,Hoa85}, are too general for
the description and analysis of the behaviours produced by sequential
programs under execution.
That is, it is quite awkward to describe and analyse behaviours of this
kind using such a general process algebra.
However, the behaviours considered in basic thread algebra can be viewed
as processes that are definable over \ACP, see e.g.~\cite{BM05c}.
This allows for the transmission protocols mentioned above to be
described and their correctness to be verified using \ACP\ or rather
\ACPt, an extension of \ACP\ which supports abstraction from internal
actions.
We consider first a very simple transmission protocol and then a more
complex one that is more efficient.

This paper is organized as follows.
First, we give brief summaries of \BTA\ (Section~\ref{sect-BTA}) and
\ACPt\ (Section~\ref{sect-ACP}).
Next, we make mathematically precise the connection between behaviours
as considered in \BTA\ and processes as considered in \ACPt\
(Section~\ref{sect-process-extr}).
After that, we describe and analyse the above-mentioned transmission
protocols (Sections~\ref{sect-protocol-1} and~\ref{sect-protocol-2}).
Finally, we make some concluding remarks (Section~\ref{sect-concl}).

\section{Thread Algebra}
\label{sect-BTA}

In this section, we review \BTA\ (Basic Thread Algebra).
\BTA\ is concerned with behaviours as exhibited by sequential programs
under execution.
These behaviours are called threads.

In \BTA, it is assumed that a fixed but arbitrary set $\BAct$ of
\emph{basic actions} has been given.
A thread performs basic actions in a sequential fashion.
Upon each basic action performed, a reply from the execution environment
of the thread determines how it proceeds.
The possible replies are the Boolean values $\True$ and~$\False$.

To build terms, \BTA\  has the following constants and operators:
\begin{iteml}
\item
the \emph{deadlock} constant $\DeadEnd$;
\item
the \emph{termination} constant $\Stop$;
\item
for each $a \in \BAct$, the binary \emph{postconditional composition}
operator $\pccop{a}$.
\end{iteml}
We assume that there are infinitely many variables, including $x,y,z$.
Terms are built as usual.
We use infix notation for the postconditional composition operator.
We introduce \emph{basic action prefixing} as an abbreviation:
$a \bapf p$, where $a \in \BAct$ and $p$ is a \BTA\ term, abbreviates
$\pcc{p}{a}{p}$.

The thread denoted by a closed term of the form $\pcc{p}{a}{q}$ will
first perform $a$, and then proceed as the thread denoted by $p$
if the reply from the execution environment is $\True$ and proceed as
the thread denoted by $q$ if the reply from the execution environment is
$\False$.
The threads denoted by $\DeadEnd$ and $\Stop$ will become inactive and
terminate, respectively.
This implies that each closed \BTA\ term denotes a thread that will
become inactive or terminate after it has performed finitely many
basic actions.
Infinite threads can be described by guarded recursion.

A \emph{guarded recursive specification} over \BTA\ is a set of
recursion equations $E = \set{X = t_X \where X \in V}$, where $V$ is a
set of variables and each $t_X$ is a \BTA\ term of the form $\DeadEnd$,
$\Stop$ or $\pcc{t}{a}{t'}$ with $t$ and $t'$ that contain only
variables from $V$.
We write $\vars(E)$ for the set of all variables that occur in $E$.
We are only interested in models of \BTA\ in which guarded recursive
specifications have unique solutions, such as the projective limit model
of \BTA\ presented in~\cite{BB03a}.

For each guarded recursive specification $E$ and each $X \in \vars(E)$,
we introduce a constant $\rec{X}{E}$ standing for the unique solution of
$E$ for $X$.
The axioms for these constants are given in Table~\ref{axioms-rec}.%
\begin{table}[!t]
\caption{Axioms for guarded recursion}
\label{axioms-rec}
\begin{eqntbl}
\begin{saxcol}
\rec{X}{E} = \rec{t_X}{E} & \mif X \!=\! t_X \in E       & \axiom{RDP}
\\
E \Implies X = \rec{X}{E} & \mif X \in \vars(E)          & \axiom{RSP}
\end{saxcol}
\end{eqntbl}
\end{table}
In this table, we write $\rec{t_X}{E}$ for $t_X$ with, for all
$Y \in \vars(E)$, all occurrences of $Y$ in $t_X$ replaced by
$\rec{Y}{E}$.
$X$, $t_X$ and $E$ stand for an arbitrary variable, an arbitrary \BTA\
term and an arbitrary guarded recursive specification over \BTA,
respectively.
Side conditions are added to restrict what $X$, $t_X$ and $E$ stand for.

In the sequel, we will make use of a version of \BTA\ in which the
following additional assumptions relating to $\BAct$ are made:
(i)~a fixed but arbitrary set $\Foci$ of \emph{foci} has been given;
(ii)~a fixed but arbitrary set $\Meth$ of \emph{methods} has been given;
(iii)~$\BAct = \set{f.m \where f \in \Foci, m \in \Meth}$.
These assumptions are based on the view that the execution environment
provides a number of services.
Performing a basic action $f.m$ is taken as making a request to the
service named $f$ to process command $m$.
As usual, we will write $\Bool$ for the set $\set{\True,\False}$.

\section{Process Algebra}
\label{sect-ACP}

In this section, we review \ACPt\ (Algebra of Communicating Processes
with abstraction).
This is the process algebra that will be used in
Section~\ref{sect-process-extr} to make precise what processes are
produced by the threads denoted by closed terms of \BTA\ with guarded
recursion.
For a comprehensive overview of \ACPt, the reader is referred
to~\cite{BW90,Fok00}.

In \ACPt, it is assumed that a fixed but arbitrary set $\Act$ of
\emph{atomic actions}, with $\tau,\dead \notin \Act$, and a fixed but
arbitrary commutative and associative function
$\funct{\commm}{\Act \x \Act}{\Act \union \set{\dead}}$
have been given.
The function $\commm$ is regarded to give the result of synchronously
performing any two atomic actions for which this is possible, and to
give $\dead$ otherwise.
In \ACPt, $\tau$ is a special atomic action, called the silent step.
The act of performing the silent step is considered unobservable.
Because it would otherwise be observable, the silent step is considered
an atomic action that cannot be performed synchronously with other
atomic actions.

\ACPt\ has the following constants and operators:
\begin{itemize}
\item
for each $e \in \Act$, the \emph{atomic action} constant $e$\,;
\item
the \emph{silent step} constant $\tau$\,;
\item
the \emph{deadlock} constant $\dead$\,;
\item
the binary \emph{alternative composition} operator $\altc$\,;
\item
the binary \emph{sequential composition} operator $\seqc$\,;
\item
the binary \emph{parallel composition} operator $\parc$\,;
\item
the binary \emph{left merge} operator $\leftm$\,;
\item
the binary \emph{communication merge} operator $\commm$\,;
\item
for each $H \subseteq \Act$, the unary \emph{encapsulation} operator
$\encap{H}$\,;
\item
for each $I \subseteq \Act$, the unary \emph{abstraction} operator
$\abstr{I}$\,.
\end{itemize}
We assume that there are infinitely many variables, including $x,y,z$.
Terms are built as usual.
We use infix notation for the binary operators.

Let $p$ and $q$ be closed \ACPt\ terms, $e \in \Act$, and
$H,I \subseteq \Act$.
Intuitively, the constants and operators to build \ACPt\ terms can be
explained as follows:
\begin{itemize}
\item
$e$ first performs atomic action $e$ and next terminates successfully;
\item
$\tau$ performs an unobservable atomic action and next terminates
successfully;
\item
$\dead$ can neither perform an atomic action nor terminate successfully;
\item
$p \altc q$ behaves either as $p$ or as $q$, but not both;
\item
$p \seqc q$ first behaves as $p$ and on successful termination of $p$
it next behaves as~$q$;
\item
$p \parc q$ behaves as the process that proceeds with $p$ and $q$ in
parallel;
\item
$p \leftm q$ behaves the same as $p \parc q$, except that it starts
with performing an atomic action of $p$;
\item
$p \commm q$ behaves the same as $p \parc q$, except that it starts with
performing an\linebreak[2] atomic action of $p$ and an atomic action of
$q$ synchronously;
\item
$\encap{H}(p)$ behaves the same as $p$, except that atomic actions from
$H$ are blocked;
\item
$\abstr{I}(p)$ behaves the same as $p$, except that atomic actions from
$I$ are turned into unobservable atomic actions.
\end{itemize}
%

The axioms of \ACPt\ are given in Table~\ref{axioms-ACPt}.
\begin{table}[!t]
\caption{Axioms of \ACPt}
\label{axioms-ACPt}
\begin{eqntbl}
\begin{axcol}
x \altc y = y \altc x                                  & \axiom{A1}  \\
(x \altc y) \altc z = x \altc (y \altc z)              & \axiom{A2}  \\
x \altc x = x                                          & \axiom{A3}  \\
(x \altc y) \seqc z = x \seqc z \altc y \seqc z        & \axiom{A4}  \\
(x \seqc y) \seqc z = x \seqc (y \seqc z)              & \axiom{A5}  \\
x \altc \dead = x                                      & \axiom{A6}  \\
\dead \seqc x = \dead                                  & \axiom{A7}  \\
{}                                                                   \\
x \parc y =
          x \leftm y \altc y \leftm x \altc x \commm y & \axiom{CM1} \\
a \leftm x = a \seqc x                                 & \axiom{CM2} \\
a \seqc x \leftm y = a \seqc (x \parc y)               & \axiom{CM3} \\
(x \altc y) \leftm z = x \leftm z \altc y \leftm z     & \axiom{CM4} \\
a \seqc x \commm b = (a \commm b) \seqc x              & \axiom{CM5} \\
a \commm b \seqc x = (a \commm b) \seqc x              & \axiom{CM6} \\
a \seqc x \commm b \seqc y =
                        (a \commm b) \seqc (x \parc y) & \axiom{CM7} \\
(x \altc y) \commm z = x \commm z \altc y \commm z     & \axiom{CM8} \\
x \commm (y \altc z) = x \commm y \altc x \commm z     & \axiom{CM9}
\end{axcol}
\qquad
\begin{axcol}
x \seqc \tau = x                                       & \axiom{B1}  \\
x \seqc (\tau \seqc (y \altc z) \altc y) = x \seqc (y \altc z)
                                                       & \axiom{B2}  \\
{}                                                                   \\
\encap{H}(a) = a                \hfill \mif a \notin H & \axiom{D1}  \\
\encap{H}(a) = \dead            \hfill \mif a \in H    & \axiom{D2}  \\
\encap{H}(x \altc y) = \encap{H}(x) \altc \encap{H}(y) & \axiom{D3}  \\
\encap{H}(x \seqc y) = \encap{H}(x) \seqc \encap{H}(y) & \axiom{D4}  \\
{}                                                                   \\
\abstr{I}(a) = a                \hfill \mif a \notin I & \axiom{TI1} \\
\abstr{I}(a) = \tau             \hfill \mif a \in I    & \axiom{TI2} \\
\abstr{I}(x \altc y) = \abstr{I}(x) \altc \abstr{I}(y) & \axiom{TI3} \\
\abstr{I}(x \seqc y) = \abstr{I}(x) \seqc \abstr{I}(y) & \axiom{TI4} \\
{}                                                                   \\
a \commm b = b \commm a                                & \axiom{C1}  \\
(a \commm b) \commm c = a \commm (b \commm c)          & \axiom{C2}  \\
\dead \commm a = \dead                                 & \axiom{C3}  \\
\tau \commm a = \dead                                  & \axiom{C4}
\end{axcol}
\end{eqntbl}
\end{table}
CM2--CM3, CM5--CM7, C1--C4, D1--D4 and TI1--TI4 are actually axiom
schemas in which $a$, $b$ and $c$ stand for arbitrary constants of
\ACPt, and $H$ and $I$ stand for arbitrary subsets of $\Act$.

A \emph{recursive specification} over \ACPt\ is a set of recursion
equations $E = \set{X = t_X \where X \in V}$, where $V$ is a set of
variables and each $t_X$ is an \ACPt\ term containing only variables
from $V$.
Let $t$ be an \ACPt\ term without occurrences of abstraction operators
containing a variable $X$.
Then an occurrence of $X$ in $t$ is \emph{guarded} if $t$ has a subterm
of the form $e \seqc t'$ where $e \in \Act$ and $t'$ is a term
containing this occurrence of $X$.
Let $E$ be a recursive specification over \ACPt.
Then $E$ is a \emph{guarded recursive specification} if, in each
equation $X = t_X \in E$:
(i)~abstraction operators do not occur in $t_X$ and
(ii)~all occurrences of variables in $t_X$ are guarded or $t_X$ can be
rewritten to such a term using the axioms of \ACPt\ in either direction
and/or the equations in $E$ except the equation $X = t_X$ from left to
right.
We only consider models of \ACPt\ in which guarded recursive
specifications have unique solutions, such as the models of \ACPt\
presented in~\cite{BW90}.

For each guarded recursive specification $E$ and each variable $X$ that
occurs in $E$, we introduce a constant $\rec{X}{E}$ standing for the
unique solution of $E$ for $X$.
The axioms for these constants are RDP and RSP given in
Table~\ref{axioms-RDP-RSP-AIP}.%
\begin{table}[!t]
\caption{RDP, RSP and AIP}
\label{axioms-RDP-RSP-AIP}
\begin{eqntbl}
\begin{saxcol}
\rec{X}{E} = \rec{t_X}{E}         & \mif X = t_X \in E  & \axiom{RDP} \\
E \Implies X = \rec{X}{E}         & \mif X \in \vars(E) & \axiom{RSP} \\
{}                                                                    \\
\multicolumn{2}{@{}l@{}}
 {\AND{n \geq 0} \proj{n}(x) = \proj{n}(y) \Implies x = y}
                                                        & \axiom{AIP}
\end{saxcol}
\qquad
\begin{axcol}
\proj{0}(a) = \dead                                     & \axiom{PR1} \\
\proj{n+1}(a) = a                                       & \axiom{PR2} \\
\proj{0}(a \seqc x) = \dead                             & \axiom{PR3} \\
\proj{n+1}(a \seqc x) = a \seqc \proj{n}(x)             & \axiom{PR4} \\
\proj{n}(x \altc y) = \proj{n}(x) \altc \proj{n}(y)     & \axiom{PR5} \\
\proj{n}(\tau) = \tau                                   & \axiom{PR6} \\
\proj{n}(\tau \seqc x) = \tau \seqc \proj{n}(x)         & \axiom{PR7}
\end{axcol}
\end{eqntbl}
\end{table}
In RDP, we write $\rec{t_X}{E}$ for $t_X$ with, for all
$Y \in \vars(E)$, all occurrences of $Y$ in $t_X$ replaced by
$\rec{Y}{E}$.
RDP and RSP are actually axiom schemas in which $X$ stands for an
arbitrary variable, $t_X$ stands for an arbitrary \ACPt\ term, and $E$
stands for an arbitrary guarded recursive specification over \ACPt.

\nopagebreak[2]
Closed terms of \ACP\ with guarded recursion that denote the same
process cannot always be proved equal by means of the axioms of \ACP\
together with RDP and RSP.
To remedy this, we introduce AIP (Approximation Induction Principle).
AIP is based on the view that two processes are identical if their
approximations up to any finite depth are identical.
The approximation up to depth $n$ of a process behaves the same as that
process, except that it cannot perform any further atomic action after
$n$ atomic actions have been performed.
AIP is given in Table~\ref{axioms-RDP-RSP-AIP}.
Here, approximation up to depth $n$ is phrased in terms of a unary
\emph{projection} operator $\proj{n}$.
The axioms for these operators are axioms PR1--PR7 in
Table~\ref{axioms-RDP-RSP-AIP}.
PR1--PR7 are actually axiom schemas in which $a$ stands for arbitrary
constants of \ACPt\ different from $\tau$ and $n$ stands for an
arbitrary natural number.

We will write $\vAltc{i \in S} p_i$, where $S = \set{i_1,\ldots,i_n}$
and $p_{i_1},\ldots,p_{i_n}$ are \ACPt\ terms,
for $p_{i_1} \altc \ldots \altc p_{i_n}$.
The convention is that $\vAltc{i \in S} p_i$ stands for $\dead$ if
$S = \emptyset$.
We will often write $X$ for $\rec{X}{E}$ if $E$ is clear from the
context.
It should be borne in mind that, in such cases, we use $X$ as a
constant.

\section{Process Extraction}
\label{sect-process-extr}

In this section, we use \ACPt\ with guarded recursion to make
mathematically precise what processes are produced by the threads
denoted by closed terms of \BTA\ with guarded recursion.

For that purpose, $\Act$ and $\commm$ are taken such that the following
conditions are satisfied:
\begin{ldispl}
\begin{aeqns}
\Act & \supseteq &
\set{\snd_f(d) \where f \in \Foci, d \in \Meth \union \Bool} \union
\set{\rcv_f(d) \where f \in \Foci, d \in \Meth \union \Bool}
\union
\set{\stp,\iact}
\end{aeqns}
\end{ldispl}%
and for all $f \in \Foci$, $d \in \Meth \union \Bool$, and
$e \in \Act$:
\begin{ldispl}
\begin{aeqns}
\snd_f(d) \commm \rcv_f(d) = \iact\;,
\\
\snd_f(d) \commm e = \dead & & \mif e \neq \rcv_f(d)\;,
\\
e \commm \rcv_f(d) = \dead & & \mif e \neq \snd_f(d)\;,
\end{aeqns}
\qquad\;
\begin{aeqns}
{} \\
\stp \commm e = \dead\;,
\\
\iact \commm e = \dead\;.
\end{aeqns}
\end{ldispl}

The \emph{process extraction} operation $\pextr{\ph}$ determines, for
each closed term $p$ of \BTA\ with guarded recursion, a closed term of
\ACPt\ with guarded recursion that denotes the process produced by the
thread denoted by $p$.
The process extraction operation $\pextr{\ph}$ is defined by
$\pextr{p} = \abstr{\set{\stp}}(\cpextr{p})$, where $\cpextr{\ph}$ is
defined by the equations given in Table~\ref{eqns-process-extr}
(for $f \in \Foci$ and $m \in \Meth$).%
\begin{table}[!t]
\caption{Defining equations for process extraction operation}
\label{eqns-process-extr}
\begin{eqntbl}
\begin{eqncol}
\cpextr{X} = X
\\
\cpextr{\Stop} = \stp
\\
\cpextr{\DeadEnd} = \iact \seqc \dead
\\
\cpextr{\pcc{t_1}{f.m}{t_2}} =
\snd_f(m) \seqc
(\rcv_f(\True) \seqc \cpextr{t_1} \altc
 \rcv_f(\False) \seqc \cpextr{t_2})
\\
\cpextr{\rec{X}{E}} =
\rec{X}{\set{Y = \cpextr{t_Y} \where Y = t_Y \,\in\, E}}
\end{eqncol}
\end{eqntbl}
\end{table}

Two atomic actions are involved in performing a basic action of the form
$f.m$: one for sending a request to process command $m$ to the service
named $f$ and another for receiving a reply from that service upon
completion of the processing.
For each closed term $p$ of \BTA\ with guarded recursion, $\cpextr{p}$
denotes a process that in the event of termination performs a special
termination action just before termination.
Abstracting from this termination action yields the process denoted by
$\pextr{p}$.
Some atomic actions introduced above are not used in the definition of
the process extraction operation for \BTA.
Those atomic actions are commonly used in the definition of the process
extraction operation for extensions of \BTA\ in which operators for
thread-service interaction occur, see e.g.~\cite{BM05c}.

Let $p$ be a closed term of \BTA\ with guarded recursion.
Then we say that $\pextr{p}$ is the \emph{process produced by} $p$.

\sloppy
The process extraction operation preserves the axioms of \BTA\ with
guarded recursion.
Roughly speaking, this means that the translations of these axioms are
derivable from the axioms of \ACPt\ with guarded recursion.
Before we make this fully precise, we have a closer look at the axioms
of \BTA\ with guarded recursion.

A proper axiom is an equation or a conditional equation.
In Table~\ref{axioms-rec}, we do not find proper axioms.
Instead of proper axioms, we find axiom schemas without side conditions
and axiom schemas with side conditions.
The axioms of \BTA\ with guarded recursion are obtained by replacing
each axiom schema by all its instances.

We define a function $\transl{\ph}$ from the set of all equations and
conditional equations of \BTA\ with guarded recursion to the set of all
equations of \ACPt\ with guarded recursion as follows:
\begin{ldispl}
\transl{t_1 = t_2} \;\;=\;\; \pextr{t_1} = \pextr{t_2}\;,
\\
\transl{E \Implies t_1 = t_2} \;\;=\;\;
\set{\pextr{t'_1} = \pextr{t'_2} \where t'_1 = t'_2 \,\in\, E} \Implies
\pextr{t_1} = \pextr{t_2}\;.
\end{ldispl}
\begin{proposition}
\label{prop-preservation-axioms}
Let $\phi$ be an axiom of \BTA\ with guarded recursion.
Then $\transl{\phi}$ is derivable from the axioms of \ACPt\ with guarded
recursion.
\end{proposition}
\begin{proof}
The proof is trivial.
\qed
\end{proof}
Proposition~\ref{prop-preservation-axioms} would go through if no
abstraction of the above-mentioned special termination action was made.
Notice further that \ACPt\ without the silent step constant and the
abstraction operator, better known as \ACP, would suffice if no
abstraction of the special termination action was made.

\section{A Simple Protocol}
\label{sect-protocol-1}

In this section, we consider a very simple transmission protocol for
passing instructions from a thread to a remote execution environment.

At the location of the thread concerned, two atomic actions are involved
in performing a basic action: one for sending a message containing the
basic action via a transmission channel to a receiver at the location of
the execution environment and another for receiving a reply via a
transmission channel from the receiver upon completion of the processing
at the location of the execution environment.
The receiver waits until a message containing a basic action can be
received.
Upon reception of a message containing a basic action $f.m$, the
receiver sends a request to process command $m$ to the service named $f$
at the location of the execution environment.
Next, the receiver waits until a reply from that service can be
received.
Upon reception of a reply, the receiver forwards the reply to the
thread.
Deadlocking and terminating are treated like performing basic actions.

We write $\BActi$ for the set $\BAct \union \set{\stopd,\deadd}$.

For the purpose of describing the very simple transmission protocol
outlined above in \ACPt, $\Act$ and $\commm$ are taken such that, in
addition to the conditions mentioned at the beginning of
Section~\ref{sect-process-extr}, the following conditions are satisfied:
\begin{ldispl}
\begin{aeqns}
\Act & \supseteq &
\set{\snd_i(d) \where i \in \set{1,2}, d \in \BActi}    \union
\set{\rcv_i(d) \where i \in \set{1,2}, d \in \BActi}
\\ & {} \union {} &
\set{\snd_i(r) \where i \in \set{3,4}, r \in \Bool} \union
\set{\rcv_i(r) \where i \in \set{3,4}, r \in \Bool} \union
\set{\jact}
\end{aeqns}
\end{ldispl}%
and for all $i \in \set{1,2}$, $j \in \set{3,4}$, $d \in \BActi$,
$r \in \Bool$, and $e \in \Act$:
\begin{ldispl}
\begin{aeqns}
\snd_i(d) \commm \rcv_i(d) = \jact \;,
\\
\snd_i(d) \commm e = \dead & & \mif e \neq \rcv_i(d)\;,
\\
e \commm \rcv_i(d) = \dead & & \mif e \neq \snd_i(d)\;,
\eqnsep
\jact \commm e = \dead\;.
\end{aeqns}
\qquad\;
\begin{aeqns}
\snd_j(r) \commm \rcv_j(r) = \jact \;,
\\
\snd_j(r) \commm e = \dead & & \mif e \neq \rcv_j(r)\;,
\\
e \commm \rcv_j(r) = \dead & & \mif e \neq \snd_j(r)\;,
\end{aeqns}
\end{ldispl}

We introduce a process extraction operation $\pextrrct{\ph}$ which
determines, for each closed term $p$ of \BTA\ with guarded recursion, a
closed term of \ACPt\ with guarded recursion that denotes the process
produced by the thread denoted by $p$ in the case where the thread is
remotely controlled.
This operation is defined by the equations given in
Table~\ref{eqns-process-extr-rct} (for $a \in \BAct$).%
\begin{table}[!t]
\caption{Process extraction for remotely controlled threads}
\label{eqns-process-extr-rct}
\begin{eqntbl}
\begin{eqncol}
\pextrrct{X} = X
\\
\pextrrct{\Stop} = \snd_1(\stopd)
\\
\pextrrct{\DeadEnd} = \snd_1(\deadd)
\\
\pextrrct{\pcc{t_1}{a}{t_2}} =
\snd_1(a) \seqc
(\rcv_4(\True) \seqc \pextrrct{t_1} \altc
 \rcv_4(\False) \seqc \pextrrct{t_2})
\\
\pextrrct{\rec{X}{E}} =
\rec{X}{\set{Y = \pextrrct{t_Y} \where Y = t_Y \,\in\, E}}
\end{eqncol}
\end{eqntbl}
\end{table}

Let $p$ be a closed term of \BTA\ with guarded recursion.
Then the process representing the remotely controlled thread $p$ is
described by
\begin{ldispl}
\encap{H}(\pextrrct{p} \parc \CHAi \parc \CHRi \parc \RCVi)\;,
\end{ldispl}
where
\begin{ldispl}
\begin{aeqns}
\CHAi & = & \hsp{.5}
\Altc{d \in \BActi}
 \rcv_1(d) \seqc \snd_2(d) \seqc \CHAi\;,
\bigeqnsep
\CHRi & = & \hsp{.7}
\Altc{r \in \Bool}
 \rcv_3(r) \seqc \snd_4(r) \seqc \CHRi\;,
\bigeqnsep
\RCVi & = &
\Altc{f.m \in \BActi}
 \rcv_2(f.m) \seqc \snd_f(m) \seqc
 (\rcv_f(\True)  \seqc \snd_3(\True) \altc
  \rcv_f(\False) \seqc \snd_3(\False)) \seqc
 \RCVi
\\ & & {} \altc
\rcv_2(\stopd) \altc \rcv_2(\deadd) \seqc \iact \seqc \dead
\end{aeqns}
\end{ldispl}
and
\begin{ldispl}
\begin{aeqns}
H & = &
\set{\snd_i(d) \where i \in \set{1,2}, d \in \BActi}    \union
\set{\rcv_i(d) \where i \in \set{1,2}, d \in \BActi}
\\ & {} \union {} &
\set{\snd_i(r) \where i \in \set{3,4}, r \in \Bool} \union
\set{\rcv_i(r) \where i \in \set{3,4}, r \in \Bool}\;.
\end{aeqns}
\end{ldispl}
$\CHAi$ is the transmission channel for messages containing basic
actions, $\CHRi$ is the transmission channel for replies, and $\RCVi$ is
the receiver.

If we abstract from all atomic actions for sending and receiving via the
transmission channels $\CHAi$ and $\CHRi$, then the processes denoted
by $\pextr{p}$ and
$\encap{H}(\pextrrct{p} \parc \CHAi \parc \CHRi \parc \RCVi)$ are equal
modulo an initial silent step.
\begin{theorem}
\label{theorem-protocol-1}
For each closed term $p$ of \BTA\ with guarded recursion:
\begin{ldispl}
\tau \seqc \pextr{p} =
\tau \seqc
\abstr{\set{\jact}}
 (\encap{H}(\pextrrct{p} \parc \CHAi \parc \CHRi \parc \RCVi))\;.
\end{ldispl}
\end{theorem}
\begin{proof}
By AIP, it is sufficient to prove that for all $n \geq 0$:
\begin{ldispl}
\proj{n}(\tau \seqc \pextr{p}) =
\proj{n}(\tau \seqc
         \abstr{\set{\jact}}
          (\encap{H}
            (\pextrrct{p} \parc \CHAi \parc \CHRi \parc \RCVi)))\;.
\end{ldispl}
This is easily proved by induction on $n$ and in the inductive step by
case distinction on the structure of $p$, using the axioms of \ACPt\ and
RDP.
\qed
\end{proof}

\section{A More Complex Protocol}
\label{sect-protocol-2}

In this section, we consider a more complex transmission protocol for
passing instructions from a thread to a remote execution environment.

The general idea of this protocol is that:
\begin{iteml}
\item
while the last basic action performed by the thread in question is
processed at the location of the receiver, the first basic actions of
the two ways in which the thread may proceed are transmitted together to
the receiver;
\item
while the choice between those two basic actions is made by the receiver
on the basis of the reply produced at the completion of the processing,
the reply is transferred to the thread.
\end{iteml}

To simplify the description of the protocol, the following extensions of
\ACP\ from~\cite{BB94a} will be used:
\begin{iteml}
\item
We will use conditionals.
The expression $\cond{p}{b}{q}$, is to be read as
\texttt{if} $b$ \texttt{then} $p$ \texttt{else} $q$.
The defining equations are
\begin{ldispl}
\cond{x}{\True}{y} = x \quad \mathrm{and} \quad
\cond{x}{\False}{y} = y\;.
\end{ldispl}
\item
We will use the generalization of restricted early input action
prefixing to process prefixing.
Restricted early input action prefixing is defined by the equation
$\erd^D_i(u) \pf t = \vAltc{d \in D} \rcv_i(d) \seqc t[d/u]$.
We use the extension to processes to express binary parallel input:
$(\erd^{D_1}_i(u_1) \parc \erd^{D_2}_j(u_2)) \pf P$.
For this particular case, we have the following equation:
\begin{ldispl}
\begin{aeqns}
(\erd^{D_1}_i(u_1) \parc \erd^{D_2}_j(u_2)) \pf t & = &
\Altc{d_1 \in D_1}
 \rcv_i(d_1) \seqc (\erd^{D_2}_j(u_2) \pf t[d_1/u_1])
\\ & & {} \altc
\Altc{d_2 \in D_2}
 \rcv_j(d_2) \seqc (\erd^{D_1}_i(u_1) \pf t[d_2/u_2])\;.
\end{aeqns}
\end{ldispl}
\end{iteml}

We write $\BActiia$ for the set $\BActi \x \BActi$, $\BActiib$ for the
set $\BAct \x \BActi \x \BActi$, and $\BActii$ for the set
$\BActiia \union \BActiib \union \set{\stopd,\deadd,\voidd}$.

For the purpose of describing the more complex transmission protocol
outlined above in \ACPt, $\Act$ and $\commm$ are taken such that, in
addition to the conditions mentioned at the beginning of
Section~\ref{sect-process-extr}, the following conditions are satisfied:
\begin{ldispl}
\begin{aeqns}
\Act & \supseteq &
\set{\snd_i(d) \where i \in \set{1,2}, d \in \BActii}    \union
\set{\rcv_i(d) \where i \in \set{1,2}, d \in \BActii}
\\ & {} \union {} &
\set{\snd_i(r) \where i \in \set{3,4}, r \in \Bool} \union
\set{\rcv_i(r) \where i \in \set{3,4}, r \in \Bool} \union
\set{\jact}
\end{aeqns}
\end{ldispl}%
and for all $i \in \set{1,2}$, $j \in \set{3,4}$, $d \in \BActii$,
$r \in \Bool$, and $e \in \Act$:
\begin{ldispl}
\begin{aeqns}
\snd_i(d) \commm \rcv_i(d) = \jact \;,
\\
\snd_i(d) \commm e = \dead & & \mif e \neq \rcv_i(d)\;,
\\
e \commm \rcv_i(d) = \dead & & \mif e \neq \snd_i(d)\;,
\eqnsep
\jact \commm e = \dead\;.
\end{aeqns}
\qquad\;
\begin{aeqns}
\snd_j(r) \commm \rcv_j(r) = \jact \;,
\\
\snd_j(r) \commm e = \dead & & \mif e \neq \rcv_j(r)\;,
\\
e \commm \rcv_j(r) = \dead & & \mif e \neq \snd_j(r)\;,
\end{aeqns}
\end{ldispl}

We introduce a process extraction operation $\apextrrct{\ph}$ which
determines, for each closed term $p$ of \BTA\ with guarded recursion, a
closed term of \ACPt\ with guarded recursion that denotes the process
produced by the thread denoted by $p$ in the case where the thread is
remotely controlled by means of the alternative transmission protocol.
This operation is defined by the equations given in
Table~\ref{eqns-process-extr-rct-alt} (for $a \in \BAct$).%
\begin{table}[!t]
\caption{Alternative process extraction for remotely controlled threads}
\label{eqns-process-extr-rct-alt}
\begin{eqntbl}
\begin{eqncol}
\apextrrct{X} = X
\\
\apextrrct{\Stop} = \snd_1(\stopd)
\\
\apextrrct{\DeadEnd} = \snd_1(\deadd)
\\
\apextrrct{\pcc{t_1}{a}{t_2}} =
\snd_1(a,\fact{t_1},\fact{t_2}) \seqc
(\rcv_4(\True) \seqc \aapextrrct{t_1} \altc
 \rcv_4(\False) \seqc \aapextrrct{t_2})
\\
\apextrrct{\rec{X}{E}} =
\rec{X}{\set{Y = \apextrrct{t_Y} \where Y = t_Y \,\in\, E}}
\eqnsep
\aapextrrct{X} = X
\\
\aapextrrct{\Stop} = \snd_1(\voidd)
\\
\aapextrrct{\DeadEnd} = \snd_1(\voidd)
\\
\aapextrrct{\pcc{t_1}{a}{t_2}} =
\snd_1(\fact{t_1},\fact{t_2}) \seqc
(\rcv_4(\True) \seqc \aapextrrct{t_1} \altc
 \rcv_4(\False) \seqc \aapextrrct{t_2})
\\
\aapextrrct{\rec{X}{E}} =
\rec{X}{\set{Y = \aapextrrct{t_Y} \where Y = t_Y \,\in\, E}}
\eqnsep
\fact{\Stop} = \stopd
\\
\fact{\DeadEnd} = \deadd
\\
\fact{\pcc{t_1}{a}{t_2}} = a
\\
\fact{\rec{X}{E}} = \fact{\rec{t_X}{E}} \quad \mif X = t_X \,\in\, E
\end{eqncol}
\end{eqntbl}
\end{table}

Let $p$ be a closed term of \BTA\ with guarded recursion.
Then the process representing the remotely controlled thread $p$ is
described by
\begin{ldispl}
\encap{H}(\apextrrct{p} \parc \CHAii \parc \CHRii \parc \RCVii)\;,
\end{ldispl}
where
\begin{ldispl}
{} \hsp{-.2}
\begin{aeqns}
\CHAii & = & \hsp{1.58}
\Altc{d \in \BActii}
 \rcv_1(d) \seqc \snd_2(d) \seqc \CHAii\;,
\bigeqnsep
\CHRii & = & \hsp{1.91}
\Altc{r \in \Bool}
 \rcv_3(r) \seqc \snd_4(r) \seqc \CHRii\;,
\bigeqnsep
\RCVii & = &
\Altc{\tup{f.m,a,a'} \in \BActiib}
 \rcv_2(f.m,a,a') \seqc \snd_f(m)
\\[-1.75ex] & & \phantom{\Altc{\tup{f.m,a,a'} \in \BActiib}} {}
\seqc
 (\rcv_f(\True)  \seqc \aRCVii(\True,a) \altc
  \rcv_f(\False) \seqc \aRCVii(\False,a'))
\\[-1.75ex] & & {} \altc
\rcv_2(\stopd) \altc \rcv_2(\deadd) \seqc \iact \seqc \dead\;,
\eqnsep
\aRCVii(r,f.m) & = &
(\snd_3(r) \parc \snd_f(m)) \seqc \aaRCVii\;,
\\
\aRCVii(r,\stopd) & = & \rcv_2(\voidd)\;,
\\
\aRCVii(r,\deadd) & = & \rcv_2(\voidd) \seqc \iact \seqc \dead\;,
\eqnsep
\aaRCVii & = &
(\erd^\BActiia_2(u,v) \parc \erd^\Bool_f(\beta)) \pf
 (\cond{\aRCVii(\beta,u)}{\beta}{\aRCVii(\beta,v)})
\end{aeqns}
\end{ldispl}
and
\begin{ldispl}
{} \hsp{-.2}
\begin{aeqns}
H & = &
\set{\snd_i(d) \where i \in \set{1,2}, d \in \BActii} \union
\set{\rcv_i(d) \where i \in \set{1,2}, d \in \BActii}
\\ & {} \union {} &
\set{\snd_i(r) \where i \in \set{3,4}, r \in \Bool} \union
\set{\rcv_i(r) \where i \in \set{3,4}, r \in \Bool}\;.
\end{aeqns}
\end{ldispl}

Notice that the first cycle of the alternative transmission protocol
differs fairly from all subsequent ones.
This difference gives rise to a slight complication in the proof of
Theorem~\ref{theorem-protocol-2} below.

If we abstract from all atomic actions for sending and receiving via the
transmission channels $\CHAii$ and $\CHRii$, then the processes denoted
by $\pextr{p}$ and
$\encap{H}(\apextrrct{p} \parc \CHAii \parc \CHRii \parc \RCVii)$ are
equal modulo an initial silent step.
\begin{theorem}
\label{theorem-protocol-2}
For each closed term $p$ of \BTA\ with guarded recursion:
\begin{ldispl}
\tau \seqc \pextr{p} =
\tau \seqc
\abstr{\set{\jact}}
 (\encap{H}(\apextrrct{p} \parc \CHAii \parc \CHRii \parc \RCVii))\;.
\end{ldispl}
\end{theorem}
\begin{proof}
By AIP, it is sufficient to prove that for all $n \geq 0$:
\begin{ldispl}
\proj{n}(\tau \seqc \pextr{p}) =
\proj{n}(\tau \seqc
         \abstr{\set{\jact}}
          (\encap{H}
            (\apextrrct{p} \parc \CHAii \parc \CHRii \parc \RCVii)))\;.
\end{ldispl}
For $n = 0,1,2$, this is easily proved.
For $n \geq 3$, it is easily proved in the cases $p \equiv \Stop$ and
$p \equiv \DeadEnd$, but in the case $p \equiv \pcc{p_1}{f.m}{p_2}$ we
get:
\begin{ldispl}
\hsp{-2.5}
\begin{geqns}
\tau \seqc \snd_f(m) \seqc
(\rcv_f(\True)  \seqc \proj{n-2}(\pextr{p_1}) \altc
 \rcv_f(\False) \seqc \proj{n-2}(\pextr{p_2}))
\\ \quad {} =
\tau \seqc \snd_f(m)
\\ \quad \phantom{{}={}} {} \seqc
(\rcv_f(\True) \seqc
 \proj{n-2}(\abstr{\set{\jact}}
             (\encap{H}
               (\aapextrrct{p_1} \parc \CHAii \parc \CHRii \parc
                \aRCVii(\True,\fact{p_1}))))
\\ \quad \phantom{{}={} {} \seqc (} {} \altc
 \rcv_f(\False) \seqc
 \proj{n-2}(\abstr{\set{\jact}}
             (\encap{H}
               (\aapextrrct{p_2} \parc \CHAii \parc \CHRii \parc
                \aRCVii(\False,\fact{p_2})))))\;.
\end{geqns}
\end{ldispl}
We have that
\begin{ldispl}
\proj{n-2}(\abstr{\set{\jact}}
            (\encap{H}
              (\aapextrrct{p'} \parc \CHAii \parc \CHRii \parc
               \aRCVii(\True,\fact{p'}))))
\\ \quad {} =
\proj{n-2}(\abstr{\set{\jact}}
            (\encap{H}
              (\aapextrrct{p'} \parc \CHAii \parc \CHRii \parc
               \aRCVii(\False,\fact{p'}))))
\end{ldispl}
in the cases $p' \equiv \Stop$ and $p' \equiv \DeadEnd$, but not in the
case $p' \equiv \pcc{p'_1}{f'.m'}{p'_2}$.
Therefore, we cannot prove
\begin{ldispl}
\proj{n}(\tau \seqc \pextr{p}) =
\proj{n}(\tau \seqc
         \abstr{\set{\jact}}
          (\encap{H}
            (\apextrrct{p} \parc \CHAii \parc \CHRii \parc \RCVii)))
\end{ldispl}
by induction on $n$.
However, in the case $p' \equiv \pcc{p'_1}{f'.m'}{p'_2}$ we have that
\begin{ldispl}
\begin{geqns}
\rcv_f(r) \seqc \proj{n-2}(\pextr{p'})
\\ \quad {} =
\rcv_f(r) \seqc \snd_{f'}(m') \seqc
\proj{n-3}(\rcv_{f'}(\True) \seqc \pextr{p'_1} \altc
           \rcv_{f'}(\False) \seqc \pextr{p'_2})
\end{geqns}
\end{ldispl}
and
\begin{ldispl}
\begin{geqns}
\rcv_f(r) \seqc
\proj{n-2}(\abstr{\set{\jact}}
            (\encap{H}
              (\aapextrrct{p'} \parc
               \CHAii \parc \CHRii \parc \aRCVii(r,f'.m'))))
\\ \quad {} =
\rcv_f(r) \seqc \snd_{f'}(m') \seqc
\proj{n-3}(\abstr{\set{\jact}}
            (\encap{H}
              (\aapextrrct{p'} \parc
               \CHAii \parc \CHRii \parc \aaRCVii)))\;.
\end{geqns}
\end{ldispl}
Therefore, it is sufficient to prove that for all closed terms $p_1$ and
$p_2$ of \BTA\ with guarded recursion, $f \in \Foci$ and $m \in \Meth$,
for all $n \geq 0$:
\begin{ldispl}
\proj{n}(\tau \seqc \
        (\rcv_f(\True) \seqc \pextr{p_1} \altc
         \rcv_f(\False) \seqc \pextr{p_2}))
\\ \quad {} =
\proj{n}(\tau \seqc
         \abstr{\set{\jact}}
          (\encap{H}
            (\aapextrrct{\pcc{p_1}{f.m}{p_2}} \parc
             \CHAii \parc \CHRii \parc \aaRCVii)))\;.
\end{ldispl}
This is easily proved by induction on $n$ and in the inductive step by
case distinction on the structure of $p_1$ and $p_2$, using the axioms
of \ACPt, RDP and the axioms concerning process prefixing and
conditionals given in~\cite{BB94a}.
\qed
\end{proof}

\section{Conclusions}
\label{sect-concl}

Using \ACPt, we have described a very simple transmission protocol for
passing instructions from a thread to a remote execution environment and
a more complex one that is more efficient, and we have verified the
correctness of these protocols.
In this way, we have clarified the phenomenon of remotely controlled
program behaviours to a certain extent.

One option for future work is to describe the protocols concerned in a
version of \ACP\ with discrete relative timing (see
e.g.~\cite{BB95a,BM02a}) and then to show that the more complex one
leads to a speed-up indeed.
Another option for future work is to devise, describe and analyse more
efficient protocols, such as protocols that allow for two or more
instructions to be processed in parallel.

By means of the protocols, we have presented a way to deal with the
instruction streams that turn up with remotely controlled program
behaviours.
By that we have ascribed a sense to the term instruction stream which
makes clear that an instruction stream is dynamic by nature, in
contradistinction with an instruction sequence.
We have not yet been able to devise a basic definition of instruction
streams.

\bibliographystyle{spmpsci}
\bibliography{TA}


\end{document}